\documentclass[]{pasj01}
\draft

\begin{document} 
\Received{}%{yyyy/mm/dd}
\Accepted{}%{yyyy/mm/dd}
%\Published{yyyy/mm/dd}

\title{Detection of New Methylamine (CH$_3$NH$_2$) Sources: Candidates for Future Glycine Surveys}

%%% begin:list of authors
% Do NOT capitalize all letters in "textsc".
\author{Masatoshi \textsc{Ohishi}\altaffilmark{1,2,3 }%
%\thanks{Example: Present Address is xxxxxxxxxx}
}
\altaffiltext{1}{National Astronomical Observatory of Japan, 2-21-1, Osawa, Mitaka, Tokyo, 181-8588 Japan}
\altaffiltext{2}{Department of Astronomical Science, SOKENDAI (the Graduate University for Advanced Studies), 2-21-1, Osawa, Mitaka, Tokyo, 181-8588 Japan}
\altaffiltext{3}{Center for Novel Science Initiatives, National Institutes of Natural Sciences, 2F Hulic Kamiyacho Building, 4-3-13 Toranomon, Minato-ku, Tokyo, 105-0001 Japan}
\email{masatoshi.ohishi@nao.ac.jp}

\author{Taiki \textsc{Suzuki}\altaffilmark{2 }
\thanks{Present address: Astrobiology Center, National Institutes of Natural Sciences, 2-21-1, Osawa, Mitaka, Tokyo, 181-8588 Japan}
}
\email{taiki.suzuki@nao.ac.jp}

\author{Tomoya \textsc{Hirota}\altaffilmark{1,2 }}
\email{tomoya.hirota@nao.ac.jp}

\author{Masao \textsc{Saito}\altaffilmark{4,2}
\thanks{Present address: National Astronomical Observatory of Japan, 2-21-1, Osawa, Mitaka, Tokyo, 181-8588 Japan}
}
\altaffiltext{4}{Nobeyama Radio Observatory, National Astronomical Observatory of Japan, Nobeyama, Minamimaki, Minamisaku, Nagano, 384-1305 Japan}
\email{masao.saito@nao.ac.jp}

\author{Norio \textsc{Kaifu}\altaffilmark{1}}
\email{norio.kaifu@nao.ac.jp}

%\author{B-Firstname \textsc{B-Familyname},\altaffilmark{2}}
%\altaffiltext{2}{B-Address of Institute}
%\email{bbbbb@xxx.xxx.xx.xx}

%\author{C-Firstname \textsc{C-Familyname}\altaffilmark{3}}
%\altaffiltext{3}{C-Address of Institute}
%\email{ccccc@xxx.xxx.xx.xx}
%%% end:list of authors

%% `\KeyWords{}' always has to be placed before `\maketitle'.
 %Do NOT move this preamble from here!

\maketitle

\begin{abstract}
It has been a long-standing problem to detect interstellar glycine (NH$_2$CH$_2$COOH), the simplest amino acid, in studying a possible relation between the Universe and origin of life.
In the last about 40 years all surveys of glycine failed, and it would be an alternative strategy to search for precursor(s) to glycine.
Such studies of precursors would be crucial prior to conducting sensitive surveys by ALMA.

Laboratory studies have suggested that CH$_3$NH$_2$ is a possible precursor to glycine.
Further theoretical study also suggested that the CH$_2$NH$_2$ radical that can be formed from CH$_3$NH$_2$ through photodissociation can be a good precursor to glycine.
Thus we observed CH$_3$NH$_2$ towards several hot core sources by using the Nobeyama 45\,m radio telescope, and succeeded in finding a new CH$_3$NH$_2$ source, G10.47+0.03, with its fractional abundance of $2.4\pm0.6\times10^{-8}$; at the time of writing, this source is the most abundant source of CH$_3$NH$_2$ ever known.
We also found another new source of CH$_3$NH$_2$, NGC6334F, however, the detection may be tentative since the signal-to-noise ratios are not sufficient to claim secure detection.
Detailed analysis of the detected data revealed the detection of the hyperfine components of CH$_3$NH$_2$ for the first time.
We found that the observed abundance of CH$_3$NH$_2$ agrees fairly well with the theoretically predicted value by \citet{Garr13}.
Detectability of interstellar glycine is discussed. 
\end{abstract}

\section{Introduction}

It is widely accepted that prebiotic chemical evolution from small to large and complex molecules would have resulted in the origin of life. 
On the other hand there are two conflicting views where inorganic formation of complex organic molecules (hereafter COMs) occurred in the early Earth, on the Earth or out of the Earth. 
\citet{Ehre02} suggested that exogenous delivery of COMs by comets and/or asteroids to the early Earth could be larger than their terrestrial formation by three orders of magnitude. 
If amino acids are formed in interstellar clouds, significant amount of them may be delivered to planets. Detection of amino acids would accelerate the discussion concerning the universality of life.

So far, many trials to detect the simplest amino acid, glycine (NH$_2$CH$_2$COOH), were made towards Sgr\,B2 and other high-mass forming regions, but none of them were successful. 
One idea to overcome this situation would be to search for precursors to glycine. 
Although the chemical evolution of interstellar N-bearing COMs is not well known, methylamine (CH$_3$NH$_2$) is proposed as one precursor to glycine. 
CH$_3$NH$_2$ can be formed from well-known abundant species, CH$_4$ and NH$_3$, on icy dust surface under cosmic ray irradiation \citep{KimK11}. 
%
%Further methyleneimine (CH$_2$NH) would be related to CH$_3$NH$_2$. 
%
Another possible route to form CH$_3$NH$_2$ is hydrogenation (addition of hydrogen) to HCN on dust surface \citep{Dick97, KimK11, Theu11}: HCN $\rightarrow$ CH$_2$NH  $\rightarrow$ CH$_3$NH$_2$. 
\citet{Theu11} made laboratory experiments of surface reactions on interstellar analog ice at 15\,K under UV irradiation.
They showed that HCN hydrogenation leads to the formation of CH$_3$NH$_2$, 
and CH$_2$NH hydrogenation leads to the formation of CH$_3$NH$_2$.
Furthermore they showed that CH$_3$NH$_2$ can thermally react with CO$_2$ in the
solid phase to form a carbamate, which can be converted in a glycine salt under VUV irradiation.

\begin{figure}[h]
 \begin{center}
  \includegraphics[width=16cm]{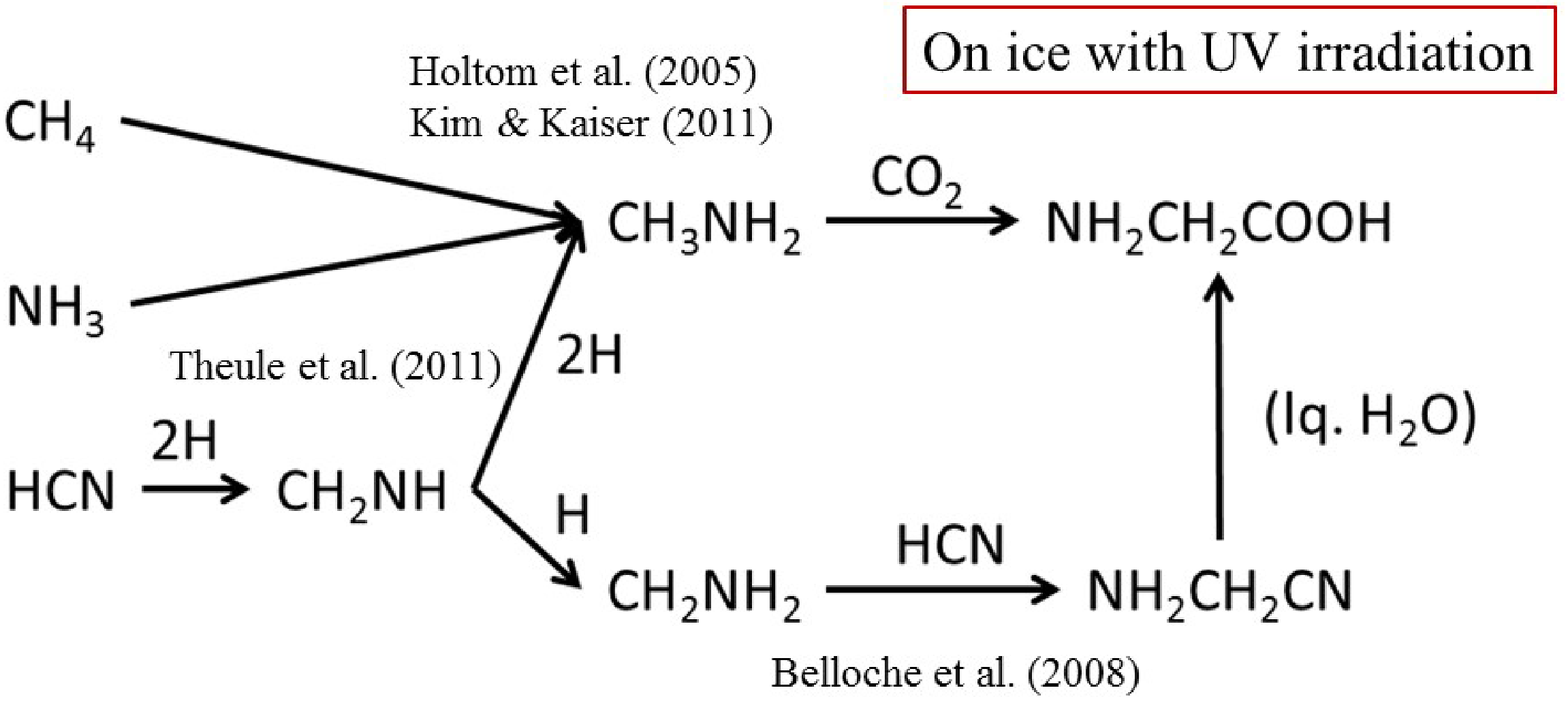} 
 \end{center}
\caption{Possible Formation Routes to Interstellar Glycine.}\label{fig:form}
\end{figure}

Figure\,\ref{fig:form} shows possible formation routes on dust surface to glycine, which is based on the laboratory studies mentioned above. 
It is well known that HCN, CH$_4$, NH$_3$ and CO$_2$ are abundant species in the interstellar molecular clouds.
When the dust is warm enough ($>20$\,K), CH$_4$ and NH$_3$ may move on the dust surface through hopping.
On the other hand when the dust is cold ($\sim10$\,K), typical for molecular clouds with no or little sign of star formation, 
the hydrogen atoms can move on the dust surface by the tunneling effect.
It is known that hydrogenation reactions to CO form CH$_3$OH in cold molecular clouds.
Since CO$_2$ exists in most of molecular clouds and CH$_3$NH$_2$ is a direct precursor candidate to glycine, CH$_3$NH$_2$-rich sources will turn into promising glycine survey targets. 
Such studies would also accelerate the discussion regarding the exogenous delivery of prebiotic species to planets and connection between the Universe and life.

Interstellar CH$_3$NH$_2$ was discovered towards Sgr\,B2 and Orion\,A \citep{Kaif74, Kaif75}.
In the 80's and 90's no dedicated studies were reported on CH$_3$NH$_2$ except for a few spectral line survey results \citep{Turn89, Numm98}.
CH$_3$NH$_2$ has recently been recognized as an important molecule since it could be 
a potential precursor to glycine \citep{Holt05, Boss09, Lee09}. 
The detection of glycine as well as CH$_3$NH$_2$ from the coma of comet 67P/Churyumov-Gerasimenko \citep{Altw16} further supports the hypothesis that CH$_3$NH$_2$ is a precursor to glycine.
\citet{Half13} observed 170 lines of CH$_3$NH$_2$ towards Sgr\,B2(N), and derived 
$T_{\rm rot}$ = $159\pm30$\,K, $N_{\rm tot}$ = $(5.0\pm0.9)\times10^{15}$\,cm$^{-2}$ 
and fractional abundance of CH$_3$NH$_2$ relative to molecular hydrogen, $f$(CH$_3$NH$_2$), of $\approx1.7\times10^{-9}$.
Astronomical significance of CH$_3$NH$_2$ has motivated to measure its millimeter-wave spectrum up to 2.6\,THz \citep{Moti14}.
There are a few observations of CH$_3$NH$_2$.
\citet{Mull11} claimed detection of CH$_3$NH$_2$ in absorption towards a quasar PKS\,1830-211, however, the signal-to-noise ratios are not sufficient to claim secure detection.
\citet{Rolf11} reported by using the SMA possible detection of two CH$_3$NH$_2$ lines towards G10.47+0.03, however, no transition information was given by the authors; a line at 354.8\,GHz (possibly the $6_1$\,B1 - $5_0$\,B2 at 354.844\,GHz) seems to be blended with a HCOOCH$_3$ line; other line would possibly be the $6_3$\,E1-1 - $5_2$\,E1-1 line at 681.324\,GHz.
\citet{Ligt15} tried to detect CH$_3$NH$_2$ in a variety of hot cores without success.
At present CH$_3$NH$_2$ has securely been known towards two sources only, and it would be crucial to find CH$_3$NH$_2$-rich sources before planning sensitive glycine surveys.

In the past CH$_2$NH was reported only in Sgr\,B2, W51, Orion\,KL, and G34.3+0.15 \citep{Dick97}. 
Thus, with a purpose to find CH$_3$NH$_2$-rich sources, in April 2013, we extended this survey by using the Nobeyama 45\,m radio telescope towards CH$_3$OH-rich sources \citep{Suzu16}. 
Our CH$_3$NH$_2$ candidate sources were selected in the following way.
Based on our working hypothesis, CH$_3$NH$_2$ would be formed through hydrogenation to HCN via CH$_2$NH.
In this case other saturated molecule, CH$_3$OH, would also be formed through hydrogenation to CO.
Since it is well known that Class\,I CH$_3$OH masers suggest high abundance of CH$_3$OH, CH$_3$OH sources may contain CH$_2$NH/CH$_3$NH$_2$ rich sources.
Thus we collected many CH$_3$OH sources from publicly available CH$_3$OH maser catalogues though Japanese Virtual Observatory portal (http://jvo.nao.ac.jp/portal/).
Then we compared these sources with those rich in other organic molecules, and selected a final set of sources to be observed.
We succeeded in detecting four new CH$_2$NH sources (G31.41+0.31, G10.47+0.03, NGC6334F, and DR21(OH)) \citep{Suzu16}. 
The derived fractional abundances of CH$_2$NH relative to H$_2$ are as high as $3\times10^{-8}$, implying that CH$_2$NH may exist widely in the ISM.
If this is the case, further hydrogenation would efficiently produce CH$_3$NH$_2$.
Therefore our next step is to investigate CH$_3$NH$_2$ towards CH$_2$NH sources in order to confirm if further hydrogenation to CH$_2$NH lead to CH$_3$NH$_2$.

In this paper we report detection of one and possibly another new CH$_3$NH$_2$ sources with high fractional abundance relative to H$_2$, then discuss possibility in detecting glycine towards these sources.

\section{Observations}
\subsection{Source Selection}
According to \citet{Suzu16} fractional abundances of CH$_2$NH relative to H$_2$ are, in descending order, G10.47+0.03, Orion\,KL, G31.41+0.31, G34.3+0.2, NGC6334F, DR\,21 and W51\,e1/e2.
All of these sources were the primary candidates to observe CH$_3$NH$_2$.
However, because of time allocation constraint and weather condition, we had to constrain ourselves to observe four sources, G10.47+0.03, G31.41+0.31, NGC6334F and W51\,e1/e2.
Table\,\ref{tab:obs} contains four sources we observed.

Here we briefly describe the four sources.

G10.47+0.03 is one of brightest hot molecular cores (HMCs) which contains four ultra-compact (UC) HII regions embedded in the hot gas traced by NH$_3$, CH$_3$CN and many other complex molecules \citep{Olmi96, Cesa98, Hatc98, Wyro99, Rolf11, Hern14}.
There is a strong 1.3\,mm continuum emission toward two of the UCHII regions.
\citet{Hern14} mapped with the SMA the central region of G10.47+0.03, and revealed that there is a very compact ($\sim0.6$\,arcsecond) and very hot (the kinetic temperature of 300-500\,K) core.
Its luminosity is $7\times10^5 L_{\solar}$ \citep{Hern14}.
The distance to G10.47+0.03 is reported to be 8.6\,kpc by measuring the trigonometric parallax \citep{Sann14}.

G31.41+0.31 is a prototypical HMC imaged in various high-excitation molecular lines of NH$_3$, CH$_3$CN, CH$_3$OH, CH$_3$CCH and other species \citep{Cesa94, Hatc98, Aray08}.
\citet{Cesa98} conducted VLA observations of highly excited NH$_3$ (4,4) line, and revealed that there exists a hot ($>200$\,K), dense ($>10^7$\,cm$^{-3}$) molecular gas within a region less than 0.1\,pc in size.
Further it is known that the source is associated with an outflow, suggesting that the source is still quite young.
The luminosity of and the distance to G31.41+0.31 are reported to be $2.5\times10^5L_{\solar}$ and 7.9\,kpc, respectively \citep{Hern14}.

NGC6334F (or NGC6334I) is the brightest far-infrared source in the northern part of NGC6334 region \citep{McBr79, Loug86} and is the dominant source at millimeter and submillimeter wavelengths \citep{Sand00}. 
It contains a Ultra Compact HII region \citep{Rodr82}, and is a site of H$_2$O, OH and CH$_3$OH masers (\cite{Elli96} and references therein). 
Further many complex organic molecules, e.g., (HCOOCH$_3$, C$_2$H$_3$CN, C$_2$H$_5$CN, CH$_3$OCH$_3$) are reported \citep{McCu00}. 
In summary NGC6334F has a complex physical structure, containing a UCHII region and hot cores \citep{Hunt06}.
The distance to NGC6334F is 1.4\,kpc by measuring the trigonometric parallax \citep{Wu14}.

W51\,e1/e2 is one of famous high-mass star forming regions with a prominent HII region.
A variety of COMs are reported detected, such as CH$_3$OH, C$_2$H$_5$OH, CH$_3$CN, (CH$_3$)$_2$O, HCOOCH$_3$ \citep{Ohis95, Dick97, Zhan98}.
Its luminosity is reported to be $1.5\times10^6L_{\solar}$ \citep{Hern14}; many recombination lines are reported \citep{Zhan98, Suzu16}. 
The distance to W51\,e1/e2 is 5.4\,kpc by measuring the trigonometric parallax \citep{Sato10}.

\subsection{Observational Procedure}
Observations were made in March and May, 2014, with the 45\,m radio telescope of the Nobeyama
Radio Observatory, National Astronomical Observatory of Japan. 
We utilized a highly sensitive superconductor-insulator-superconductor (SIS) mixer receiver, TZ, in a frequency range between 78 and 100\,GHz in the dual-polarization mode.
The system temperature for each polarization was 120--200\,K during our observations.
At 86\,GHz (in the middle of the observed frequency range), the main beam efficiency ($\eta_{\rm mb}$) was about 0.4, and the beam size (FWHM) was about 20".
The pointing accuracy was checked every about two hours by observing SiO masers ($v$=1 and 2, $J$=1-0) towards red giant stars near observed sources. 
The pointing error was estimated to be within 5''.
Digital spectrometers (SAM45) with the channel spacing of 122\,kHz were used for the backend.
The on-source integration time per line per source was {2 400}--{3 600} seconds.
The individual polarization data were reduced separately, which were then co-added to generate final data with much better signal-no-noise ratio.

Table\,\ref{tab:lines} summarizes the calculated rest frequencies in MHz, the energy of the lower state ($E_l$) in Kelvin, the product $\mu^2S$ of the square of the dipole moment $\mu^2$ and transition line strength ($S$) of observed CH$_3$NH$_2$ lines \citep{Ilyu07}.
It is noted that, in Table\,\ref{tab:lines}, the frequencies without quantum number $F$ are those when the hyperfine structure (hfs) splittings due to the nuclear spin angular momentum for the $^{14}{\rm N}$ nucleus are not considered.
The intensities corresponding to such lines are the sum of those with the hfs splittings.

\section{Results}
\subsection{New CH$_3$NH$_2$ Sources}

\subsubsection{G10.47+0.03}

\begin{figure}[h]
 \begin{center}
  \includegraphics[width=18cm]{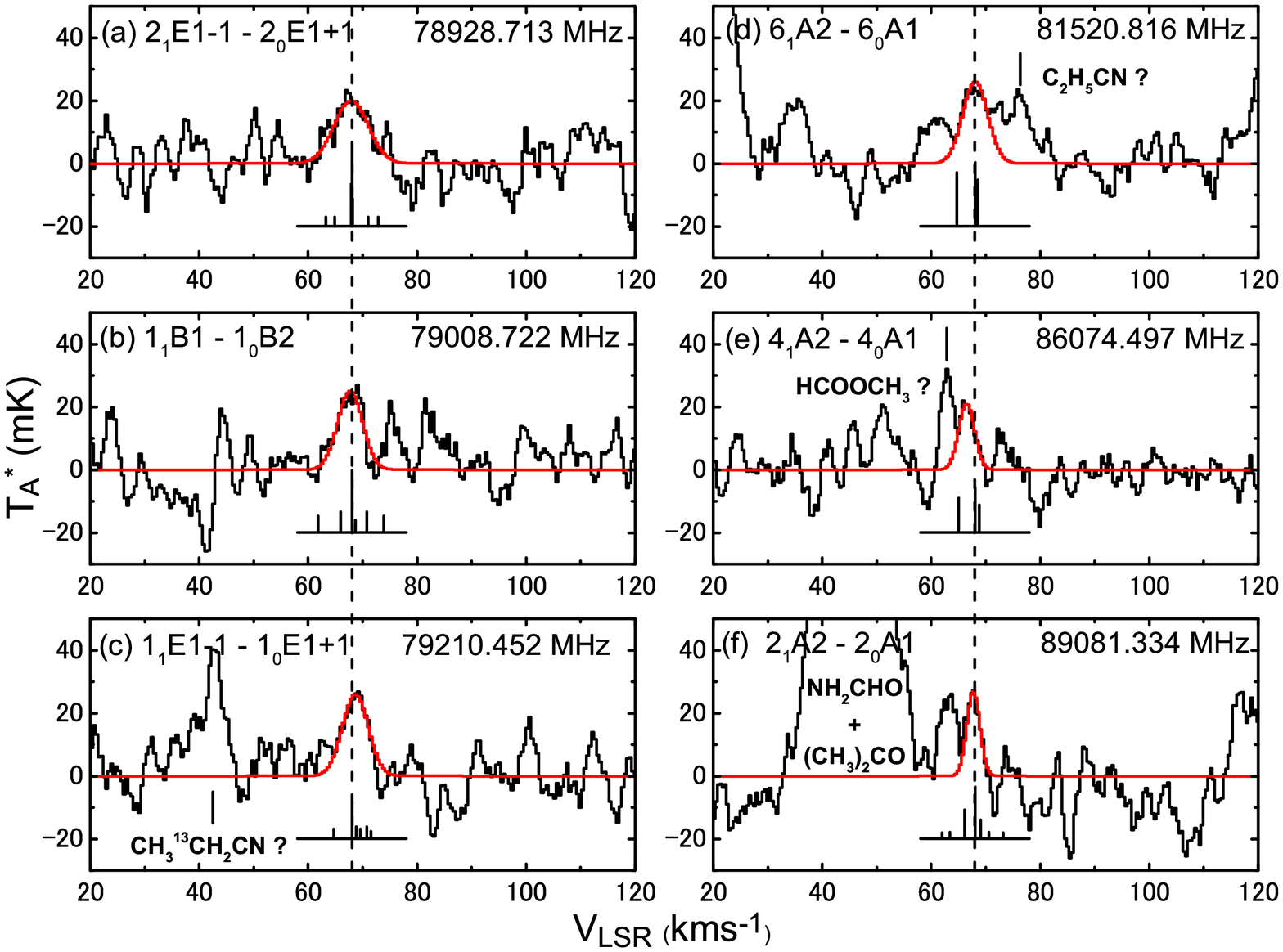} 
 \end{center}
\caption{CH$_3$NH$_2$ spectra detected towards G10.47+0.03. The abscissa is the radial velocity, 
and the vertical dashed lines correspond to the systemic radial velocity of G10.47+0.03, 68\,km\,s$^{-1}$. The red curves show results of the Gaussian fitting to the observed spectra.
On each panel the transition is shown on the top-left, and the reference frequency which was used in calculating the radial velocity is shown on the top-right. We adopted the calculated transition frequencies correspinding to the strongest hyperfine component, taken from \citet{Ilyu07}, as the reference frequencies. We also indicate positions and relative intensities (they do not scale among the six panels) of the hfs compoments. We have omitted to describe the $F$ quantum numbers so as not to make each panel too crowded. See Table\,\ref{tab:lines} for identifiying each hyperfine component line.}\label{fig:G10}
\end{figure}

Figure\,\ref{fig:G10} shows observed spectra towards G10.47+0.03.
The Gaussian-fit lines are also shown in red solid lines, where the Gaussian-fit parameters are summarized in Table\,\ref{tab:gauss_G10}.
In the initial identification process, we adopted rest frequencies without considering hfs splittings.
Then we found some lines of CH$_3$NH$_2$ showed small deviations from the source systemic radial velocity, 68\,km\,s$^{-1}$.
Thus we adopted the transition frequency of the strongest hfs component for each transition as the rest frequency.
Then we found more consistent radial velocities among the CH$_3$NH$_2$ line.

In Figures\,\ref{fig:G10}(a), (b) and (c), three lower frequency CH$_3$NH$_2$ lines (78\,929, 79\,008 and 79\,210 MHz) are clearly seen despite the signal-to-ratio of $\sim$3. 
In Table\,\ref{tab:gauss_G10} the line at 78\,929\,MHz shows slightly broader linewidth than other ones.
This transition, 2$_1$\,E1-1 - 2$_0$\,E1+1, has seven hfs components between 78\,927.460 and 78\,929.991\,MHz (see Table\,\ref{tab:lines}); three hfs components ($F$=1-1, 3-3 and 2-2) dominate among the seven components.
The broader linewidth would be reconciled by overlapping of these hfs components, primarily by the strongest three lines.
We calculated line width corresponding to the ``overlapped'' line; 
for a typical line width of 6\,km\,s$^{-1}$ in G10.47+0.03 the apparent linewidth of the ``overlapped'' line becomes to be about 7.4\,km\,s$^{-1}$, which is very close to the measured value.
The peak radial velocity for this transition is measured to be 68.1\,km\,s$^{-1}$ which is consistent with the source systemic radial velocity of 68\,km\,s$^{-1}$.
The lines at 79\,008.722\,MHz (Figure\,\ref{fig:G10}(b)) and 79\,210.452\,MHz (Figure\,\ref{fig:G10}(c)) also have hyperfine splittings, however, only a single hfs component dominates, respectively (see Table\,\ref{tab:lines}).
As are seen in Table\,\ref{tab:lines}, the $F$=2-2 component of the 1$_1$\,B1 - 1$_0$\,B2 transition at 79\,008.693\,MHz has the line intensity ($\mu^2S$) of  0.989, which is stronger than other hfs components by more than a factor of 3.
Similarly, the $F$=2-2 component of the 1$_1$\,E1 - 1-1$_0$\,E1+1 transition at 79\,210.297\,MHz has the line intensity ($\mu^2S$) of  0.580, which is stronger than other hfs components by more than a factor of 3.
Therefore we may regard that only the strongest hfs components have been detected for these transitions.
The measured radial velocities are, respectively, 68.3 and 67.3\,km\,s$^{-1}$, consistent with the source systemic radial velocity.
The line widths are 5.4 and 5.6\,km\,s$^{-1}$, which are also consistent with the typical value of 6\,km\,s$^{-1}$ in G10.47+0.03.
The peak radial velocities and line widths for the three transitions (2$_1$\,E1-1 - 2$_0$\,E1+1, 1$_1$\,B1 - 1$_0$\,B2 and 1$_1$\,E1-1 - 1$_0$\,E1+1) agree fairly well, indicating that they originate from a single common region.
In Figure\,\ref{fig:G10}(c), there is a peak at at $V_{\rm LSR}\sim43$\,km\,s$^{-1}$.
The line may be assigned to the 9$_{09}$ - 8$_{08}$ transition of CH$_{3}^{13}$CH$_{2}$CN whose lower energy level of 15\,K.

The frequency region including the 6$_1$\,A2 - 6$_0$\,A1 line with its rest frequency without hfs splittings of 81521.078\,MHz shows a ``mountain-like'' complicated feature.
The feature was decomposed into three gaussian lines with their peak frequencies of 81\,518.555, 81\,520.789 and 81\,522.812\,MHz.
The second one would correspond to the 6$_1$\,A2 - 6$_0$\,A1 line (Figure\,\ref{fig:G10}(d)) since its radial velocity was measured to be 67.9\,km\,s$^{-1}$ for the assumed rest frequency of 81\,520.816\,MHz.
Indeed the $F$=7-7 and  the $F$=5-5 components are so close in frequency (154\,kHz), which is within the frequency resolution of 244\,kHz (twice the channel spacings).
On the other hand, the $F$=6-6 component is separated from the $F$=7-7 component by 915\,kHz.
Further the measured line width after the gaussian fittings was 5.4\,km\,s$^{-1}$ which is close to those of the lines at 79\,008.722 and 79\,210.452\,MHz.
Therefore we may consider that the line at 81\,520.816\,MHz would be an overlapped line of the $F$=7-7 and 5-5 components.
The $F$=6-6 component has about a half of the line intensity (3.438) than the overlapped hfs components (7.046); the $F$=6-6 component may be buried in the unidentified line at 
81\,522.812\,MHz (at $V_{\rm LSR}\sim62$\,km\,s$^{-1}$).
The 81\,518.555\,MHz line (at $V_{\rm LSR}\sim76.3$\,km\,s$^{-1}$) may be the $8_{4,4}-9_{3,7}$ transition of C$_2$H$_5$CN at 81\,518.050\,MHz.
For this line another possibility is the $N$=8-7, $J$=17/2-15/2 transition of MgCN at 
81\,518.677\,MHz, however, MgCN radical has so far been detected only towards the circumstellar shell of IRC+10126.
Many C$_2$H$_5$CN lines have been detected towards G10.47+0.07, and it would be more plausible that the 81\,518.555\,MHz line is due to the $b$-type transition of C$_2$H$_5$CN.

We also tentatively detected two other lines, 86\,074 and 89\,081\,MHz.
The 4$_1$\,A2 - 4$_0$\,A1 line (Figure\,\ref{fig:G10}(e)) shows the radial velocity of 69.5\,km\,s$^{-1}$, which would be acceptable as a molecular line from G10.47+0.03.
However, its intensity is not enough and the line width is narrower (3.4\,km\,s$^{-1}$) than the definitely detected four lines.
The line next to the possible 4$_1$\,A2 - 4$_0$\,A1 line (at $V_{\rm LSR}\sim63$\,km\,s$^{-1}$; its frequency is 86\,075.977\,MHz) could be a methyl formate (HCOOCH$_3$) line in its first torsionally excited state. 
The 2$_1$\,A2 - 2$_0$\,A1 line (Figure\,\ref{fig:G10}(f)) shows the radial velocity of 68.2\,km\,s$^{-1}$, consistent with the four detected CH$_3$NH$_2$ lines above.
However, the measured line width (2.8\,km\,s$^{-1}$) is so narrow as a molecular line from G10.47+0.03.
This could be due to the relatively higher noise level at around the observed frequency range.

Thus we are able to conclude that we have successfully detected four lines and possibly two additional lines of CH$_3$NH$_2$ towards G10.47+0.03.
At the best of our knowledge this is the first secure report on detection of CH$_3$NH$_2$ towards G10.47+0.03.
\vspace{3mm}

\subsubsection{NGC6334F}
\begin{figure}[h]
 \begin{center}
  \includegraphics[width=18cm]{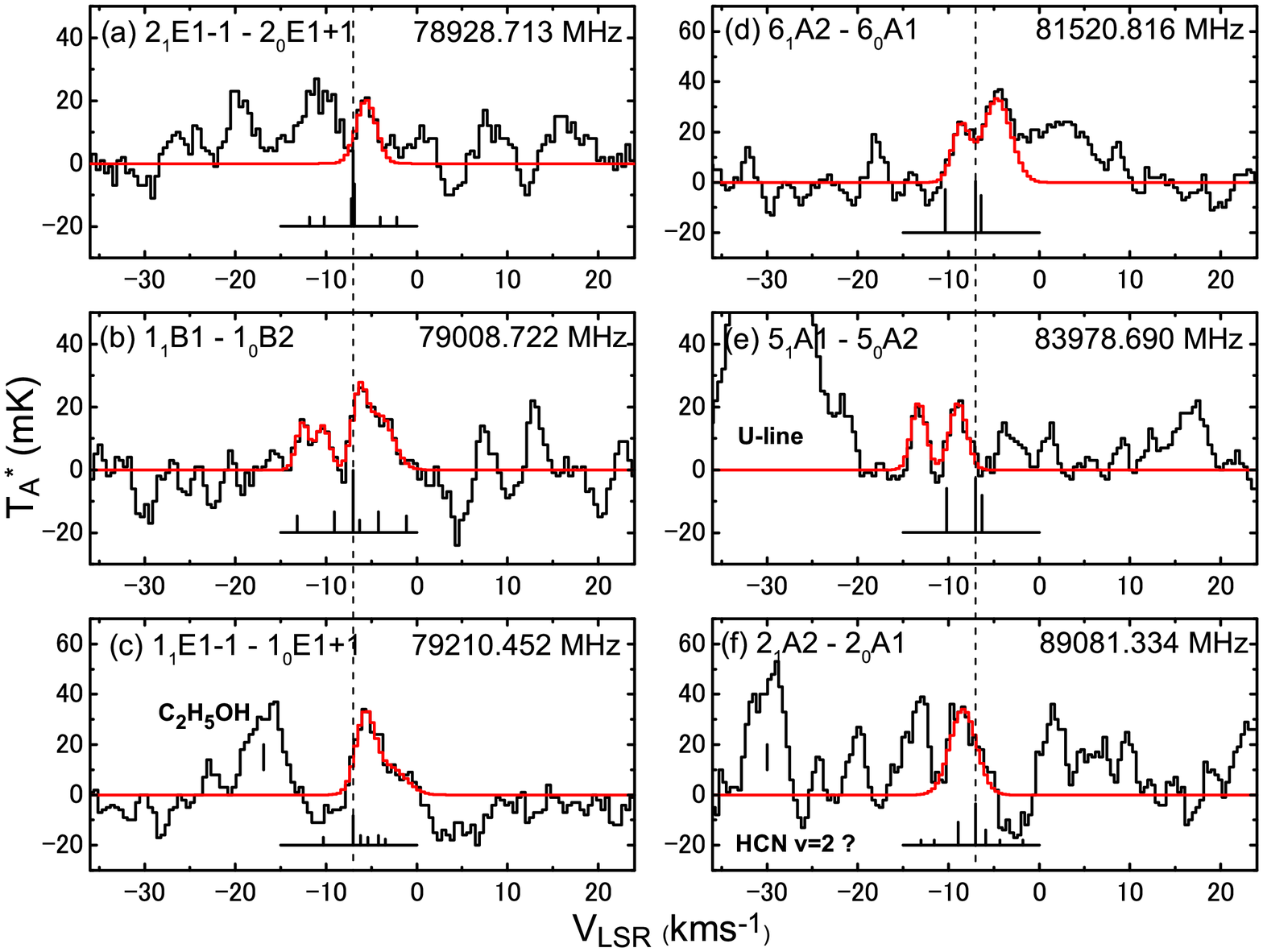} 
 \end{center}
\caption{CH$_3$NH$_2$ spectra detected towards NGC6334F. The abscissa is the radial velocity, 
and the vertical dashed lines correspond to the systemic radial velocity of NGC6334F, -7\,km\,s$^{-1}$. The red curves show results of the Gaussian fitting to the observed spectra. 
On each panel the transition is shown on the top-left, and the reference frequency which was used in calculating the radial velocity is shown on the top-right. We adopted the calculated transition frequencies correspinding to the strongest hyperfine component, taken from \citet{Ilyu07}, as the reference frequencies. We also indicate positions and relative intensities (they do not scale among the six panels) of the hfs compoments. We have omitted to describe the $F$ quantum numbers so as not to make each panel too crowded. See Table\,\ref{tab:lines} for identifiying each hyperfine component line.
It is noted that the displayed velocity range is 60\,km\,s$^{-1}$, which is a half of the case for G10.47+0.03 (Figure\,\ref{fig:G10}).}\label{fig:N63}
\end{figure}

Figures\,\ref{fig:N63}(a) through (f) show observed spectra towards NGC6334F.
The Gaussian-fit lines are also shown in red solid lines, where the Gaussian-fit parameters are summarized in Table\,\ref{tab:gauss_N6334}.
At the first glance we found many peaks in the spectra.
While the initial line identification process, we referred to the rest frequencies without considering the hyperfine splittings (similar to the case for G10.47+0.03).
In this case candidate peaks for CH$_3$NH$_2$ lines had very scattered radial velocities between about 5 and 10\,km\,s$^{-1}$.
Since a typical line width observed towards NGC6334F is around 2\,km\,s$^{-1}$, which is comparable to the hyperfine splittings for some CH$_3$NH$_2$ lines (see Table\,\ref{tab:lines}), we investigated a possibility that the hyperfine structure lines are observed (partially) resolved.
Then we were able to find hints of several hfs lines of the CH$_3$NH$_2$; they have more consistent radial velocity distribution among the lines.
The result is summarized in Table\,\ref{tab:gauss_N6334}.

However we need to consider that the measured radial velocities for these peaks still deviated to some extent from the source systemic radial velocity of NGC6334F, -7\,km\,s$^{-1}$.
One and plausible possibility would be that the observed data are too noisy and the ''true'' peaks are distorted by the poor signal-to-noise ratios.
In this case we need to be very careful since we might have made misidentifications especially for very weak peaks.
The second and less plausible possibility is that the transition frequencies of the hfs components have large uncertainties.
The hfs frequencies we adopted are those theoretically calculated ones, not actually measured in laboratories.
\citet{Taka73} measured frequencies of some hfs components of CH$_3$NH$_2$ $J < 6$ in laboratory; the reported hfs frequencies agree with the radioastronomically measured ones in this work within measurement uncertainty.
\citet{Taka73} stated that the uncertainty of the laboratory measurement is less than 100\,kHz for most lines, but it may be larger for those lines which showed no resolved hyperfine structure; the unresolved one is, e.g., the 2$_1$\,A2 - 2$_0$\,A1 transition at 89\,081\,MHz.
In this regard, it may be needed to reconsider the analysis results made by \citet{Ilyu07}.
Since the frequency measurements by the radio astronomical techniques sometimes give much better accuracy than the cases in laboratories, it would be necessary, in the future, to conduct collaborative study for improving the hyperfine structure constants between radio astronomers and laboratory microwave spectroscopists.

The 1$_1$\,B1 - 1$_0$\,B2 transition at about 79\,008\,MHz (Figure\,\ref{fig:N63}(b)) showed a blend of three hfs components ($F$=1-1, 2-2 and 2-1) and possibly two more hfs components ($F$=1-2 and 1-0).
The gaussian-fitting result showed that there are apparently two ``lines''; one would correspond to unresolved hfs lines ($F$=1-1 and 2-2; at $V_{\rm LSR}=-6.3$\,km\,s$^{-1}$) and another one ($F$=2-1; at $V_{\rm LSR}=-3.3$\,km\,s$^{-1}$ in Figure\,\ref{fig:N63}(b)) would partially overlapped with the $F$=1-1 and 2-2 component.
These two lines show consistent radial velocities (-6.7 and -6.3\,km\,s$^{-1}$), which are close to the typical radial velocity for NGC6334F.
In this case their theoretical intensity ratio would be (0.989+0.198):0.330 = 1.187:0.330$\approx$3.6.
However the observed intensity ratio is 23:17$\approx$1.4, which does not match with the theoretical one.
The difference would be due to the noise around the hfs lines.
Much weaker components ($F$=1-2 and 1-0) have lower signal-to-noise ratios of less than three.
Because these two lines would strongly be affected by the noise, we think that these lines are tentatively detected hfs components.

The 1$_1$\,E1-1 - 1$_0$\,E1+1 transition at about 79\,210\,MHz may also show hfs components (Figure\,\ref{fig:N63}(c)).
A blend of the $F$=1-2 and the $F$=2-2 components peaks at 79\,210\,MHz with its radial velocity of -5.7\,km\,s$^{-1}$.
A weaker component is seen at the pedestal of the blend of the $F$=1-2 and the $F$=2-2 components, which can be identified to be the $F$=2-1 component.
In Figure\,\ref{fig:N63}(c) the peak at a velocity of about -16\,km\,s$^{-1}$ has the measured frequency of 79\,213\,MHz, and would be assigned to the $14_{4,10}-14_{3,12}$ g$^{+}$ transition of C$_{2}$H$_{5}$OH with its calculated frequency of 79\,213.385\,MHz.
The 6$_1$\,A2 - 6$_0$\,A1 transition at about 81\,520\,MHz (Figure\,\ref{fig:N63}(d)) show two peaks which would correspond to a blend of $F$=5-5 and the $F$=7-7 components and the $F$=6-6 component.
Since the frequency separation between the $F$=5-5 and the $F$=7-7 components is 154\,kHz which is similar to the channel spacing of 122\,kHz, these two components can not be resolved.
The former line would have summed intensity of 2.968+4.078=7.046, and the latter has 3.438 (see Table\,\ref{tab:gauss_N6334}).
Thus the intensity ratio would be 7.046:3.438$\approx$2; the observed intensity ratio is 33:23$\approx$1.4.
These ratios would not be inconsistent when we take the noise level into account.
The observed radial velocity for the $F$=6-6 component is -5.2\,km\,s$^{-1}$.
If we assume a rest frequency of 81\,520.74\,MHz, which is an average between the $F$=5-5 and the $F$=7-7 components, the corresponding radial velocity becomes -4.9\,km\,s$^{-1}$, which would not be inconsistent with that of the $F$=6-6 component.
We should note that the deviation of the radial velocities of these lines from the source systemic velocity are around 2.5\,km\,s$^{-1}$, which are relatively large.

For the 5$_1$\,A1 - 5$_0$\,A2 transition at about 83\,979\,MHz (Figure\,\ref{fig:N63}(e)), two peaks corresponding to a blend of the $F$=6-6 and $F$=4-4 components and the $F$=5-5 component.
The measured apparent radial velocity of the blended one, -8.9\,km\,s$^{-1}$, would be acceptable, but the radial velocity of the $F$=5-5 component, -10\,km\,s$^{-1}$, would be too much deviated from the source systemic velocity.
On this panel there is a strong line (peak antenna temperature of 254\,mK) at a velocity of $\sim$-30\,km\,s$^{-1}$; its measured frequency is 83\,985.141\,MHz and the linewidth is 6.0\,km\,s$^{-1}$, much broader than a typical line towards NGC6334F.
Unfortunately we were not able to find any candidates for this line in Splatalogue and a list of radio recombination lines, and the line is remained unidentified.
For the 2$_1$\,A2 - 2$_0$\,A1 transition at about 89\,081\,MHz (Figure\,\ref{fig:N63}(f)), only the strongest $F$=3-3 component would be seen with the radial velocity of -8.4\,km\,s$^{-1}$.
There is a line at a velocity of $\sim$-29\,km\,s$^{-1}$ in Figure\,\ref{fig:N63}(f).
Its peak frequency, after the gaussian-fitting, is measured to be 89\,088.141\,MHz.
We may attribute the line to the $v=2$ $J=1-0$ $l=0$ transition of HCN at 89\,087.917\,MHz, however, its energy level exceeds 2\,000\,K.
Thus it would be premature to conclude its secure detection, although the detection had been reported \citep{Luca88}.
And for the 2$_1$\,E1-1 - 2$_0$\,E1+1 transition at about 78\,928\,MHz (Figure\,\ref{fig:N63}(a)), there is a hint for the $F$=3-3 component, however, the signal-to-noise ratio is too poor to claim a detection.

In summary it is possible that CH$_3$NH$_2$ is detected towards NGC6334F, with some hyperfine components, however, we dare to say that detection is not so secure due to insufficient signal-to-noise ratios.
\vspace{3mm}

\subsection{Non-Detections}
We were not able to securely detect CH$_3$NH$_2$ towards other two sources, G31.41+0.31 and W51\,e1/e2.
For G31.41+0.31 we recognized some hints of CH$_3$NH$_2$, with an rms noise level of 8\,mK, but the signal-to-noise ratios were too poor to claim new detection.
This result is consistent with the non-detection report of CH$_3$NH$_2$ by \citet{Ligt15}.
For W51\,e1/e2, we were not able to see any hint of CH$_3$NH$_2$, with an rms noise level of 8\,mK.
\vspace{3mm}

\subsection{Estimation of Excitation Temperatures and Column Densities of CH$_3$NH$_2$}
For the newly detected sources (G10.47+0.03 and NGC6334F), there are more than two lines each.
Since the lines are so weak, the detected lines can be regarded optically thin.
Thus we employed the rotation diagram method in deriving the excitation temperatures and the column densities of CH$_3$NH$_2$.

In this case we need to take the source coupling factors (${\theta}_{\rm s}$/20)$^{2}$, where ${\theta}_{\rm s}$ is the source size in arcsecond) into account for deriving reliable excitation temperatures and column densities.
Since we observed a single position for each of our sources, the source coupling factors are hard to know prior to our analyses.
The observed frequencies are between 79 and 89\,GHz, source coupling factors may not greatly vary from transition to transition, if the source size is constant for each source.
The beam sizes vary only about 1 arcsecond between 79 and 89\,GHz, i.e., 5\,\% relative to 20''.
Thus the change of the beam filling factor is only 10\% (=$1.05^{2}-1$), and we may assume a single constant beam filling factor.
In this case the excitation temperatures may not be affected by an assumed source coupling factor; only the column densities would vary by a constant factor.
Thus we first assume that the sources fill the beam for the both sources (source coupling factor of unity); the derived column densities are {\it beam-averaged column densities} which correspond to the lower limits to the column densities.
And, if we found the derived individual excitation temperature suggests that the source is compact, we need to derive {\it source-averaged column density} through dividing the {\it beam-averaged column density} by the source coupling factor.

The derived excitation temperatures are 64$\pm$46 and 37$\pm$20\,K for G10.47+0.03 and NGC6334F, respectively.
These values are sufficiently low compared with the reported excitation temperature for Sgr\,B2(N), $159\pm30$\,K \citep{Half13}.
For G10.47+0.03 \citet{Hofn00} observed the the $J$=1-0 and 2-1 rotational transitions of
C$^{17}$O and the $J$=2-1 transition of C$^{18}$O with the IRAM\,30\,m telescope.
They derived the rotational temperature of 16\,K, and showed that the C$^{17}$O emission extends to a 53"$\times$45" region.
Further, \citet{Wien12} observed G10.47+0.03 with the NH$_3$ (1,1), (2,2) and (3,3) lines by the Effelsberg 100\,m telescope with the beam size of 40", and derived the kinetic temperature towards the cold high mass clump to be $27\pm2$\,K.
Considering the estimated uncertainty of the rotational excitation temperature of CH$_3$NH$_2$, it would be slightly higher than that of C$^{17}$O; the source coupling factor may be smaller than unity.

\citet{Font07} derived excitation temperatures of three organic molecules (C$_2$H$_5$CN, C$_2$H$_3$CN and CH$_3$OCH$_3$) towards G10.47+0.03 by using the IRAM 30\,m telescope, which are $103\pm12$, $176\pm35$ and $156\pm37$\,K, respectively.
They assumed the source size of 1.3'' \citep{Olmi96}.
\citet{Hern14} mapped G10.47+0.03 by the SMA in several lines of CH$_3$CN, and reported that the CH$_3$CN emission consists of two components: a compact source (the source size of 0.6'' and the excitation temperature of 408\,K) and an extended source (4'' and 82\,K).
Based on these reported source sizes we may infer that the source size for CH$_3$NH$_2$ would be larger than 4''.
%
%\citet{Suzu17} argued the source size of G10.47+0.03, and suggested that a source size of 10'' would be applicable.
%
It is generally known that low-lying transitions tend to have larger source sizes compared with those of high-lying transitions.
Since our observed CH$_3$NH$_2$ have energy levels lower than 45\,K (see Table\,\ref{tab:lines}) and the critical densities of our observed transitions ($4\times10^{4}-1\times10^{5}$\,cm$^{-3}$) are similar within a factor of about 2 , in this paper, we assume that the source size is 10'' (a linear size of 0.4\,pc) .
The corresponding source coupling factor is 0.25.

The beam-averaged column density of CH$_3$NH$_2$ derived by the rotational diagram method is $1.6\pm0.4\times10^{15}$\,cm$^{-2}$.
Then the source-averaged column density of CH$_3$NH$_2$ is $6.3\pm1.6\times10^{15}$\,cm$^{-2}$.
With the column density of H$_2$ ($2.6\times10^{23}$\,cm$^{-2}$), derived from the C$^{17}$O observations by \citet{Half13}, the column density of CH$_3$NH$_2$ can be converted into the fractional abundance to H$_2$ to be $2.4\pm0.6\times10^{-8}$.

The situation is similar to the case for NGC6334F.
\citet{McCu00} conducted CO and CS observations towards NGC6334F by using the ACTA, and reported that the CO emission originates in an extended (25-30 arcseconds), relatively low temperature ($T_{\rm kin}$=44\,K) region.
Since the gas kinetic temperature above is close to the excitation temperature derived by the rotation diagram method, we may assume that the CH$_3$NH$_2$ emission fills the beam.
Therefore we may adopt the excitation temperature of 37$\pm$20\,K for NGC6334F.
The corresponding source- (also beam-) averaged column density of CH$_3$NH$_2$ is $6.8\pm2.5\times10^{14}$\,cm$^{-2}$.
Similar to the case for G10.47+0.03, the column density of CH$_3$NH$_2$ can be converted into the fractional abundance to H$_2$ to be $4.9\pm1.8\times10^{-10}$.

We note that the derived fractional abundance of CH$_3$NH$_2$, $2.4\pm0.6\times10^{-8}$, is higher than that in Sgr\,B2 (1.7$\times10^{-9}$; \citet{Half13}) about an order of magnitude.
In other words G10.47+0.03 is the most CH$_3$NH$_2$-rich source ever known.

We calculated the upper limits to the column densities of CH$_3$NH$_2$ for G31.41+0.31 and W51\,e1/e2, based on the 1$_1$\,B1 - 1$_0$\,B2 line at 79\,008\,MHz.
We assumed an excitation temperature between 37 and 64\,K by referring to the cases of G10.47+0.03 and NGC6334F, and adopted the same linewidths as those of CH$_2$NH observed towards G31.41+0.31 and W51\,e1/e2 (2.4 and 11.6\,km\,s$^{-1}$, respectively \citep{Suzu16}) .
The obtained upper limits to the column densities of CH$_3$NH$_2$ for G31.41+0.31 and W51\,e1/e2 are 2.4--5.3 and 12--26 $\times10^{14}$\,cm$^{-2}$, respectively.

\section{Discussion}
\subsection{Comparison with Theoretical Studies}
\citet{Garr13} conducted simulation studies that incorporated gas-phase, grain-surface, and bulk-ice chemistry in hot cores.
He constructed a chemical network to investigate formation mechanisms to glycine and related molecules such as glycinal, and found that peak gas-phase glycine fractional abundances are in the range $8\times10^{-11}$--$8\times10^{-9}$, depending on adopted warm-up models (fast, medium and slow warm-up cases).
Although he was not able to identify a single dominant mechanism to form glycine, he suggested that a reaction between two radicals (CH$_2$NH$_2$ and HOCO) would be the major contributor to form glycine.
\citet{Suzu17} investigated possible formation paths to glycine more than those studied by \citet{Garr13}, and confirmed that the path CH$_2$NH$_2$ + HOCO $\rightarrow$ NH$_2$CH$_2$COOH is the most efficient one.

The CH$_2$NH$_2$ radical can be formed on grain surfaces by photo dissociation of CH$_3$NH$_2$:
CH$_3$NH$_2$ + h$\nu$ $\rightarrow$ CH$_2$NH$_2$ + H .
Then the hydrogen atom would react with CO$_2$ which is known to be rich and ubiquitous on grains to form the HOCO radical: CO$_2$ + H $\rightarrow$ HOCO .
Although \citet{Garr13} suggested that the HOCO radical can be formed from HCOOH (formic acid), not from CO$_2$, the final reaction to form glycine was identical to the above argument.

Thus it would be worthy to compare our observational results and theoretical simulation results by \citet{Garr13} to see if the simulation can reasonably reconcile the observational results.
We summarize the comparison results in Table\,\ref{tab:comparison}.
In this Table we also included the observational results on CH$_2$NH obtained by \citet{Suzu16}.
It is seen that the medium warm-up model best matches with the observed values for the case of G10.47+0.03 if we allow a tolerance of a factor of 3.
If we consider the fact that a tolerance of a factor of 5 has been usually used in comparing simulated and observed values, the agreement is very good.
On the other hand the observed values for NGC6334F are much lower than the simulated values.
It means that the physical condition in NGC6334F would be different from the peak condition used in the simulation.

\subsection{Possibility to Detect Glycine toward G10.47+0.03}
We have seen a good agreement between our observed values and simulated ones by \citet{Garr13}.
\citet{Garr13} argued detectability of glycine towards NGC6334 IRS1, which is different from NGC6334F, and claimed that it would be possible to detect glycine using ALMA with 32 antennas (Cycle\,1) in one hour of on-source integration time.
Now ALMA has equipped with 50 12\,m antennas, and its sensitivity is much higher than that in its Cycle 1.
In this subsection we will discuss the detectability of glycine towards G10.47+0.03 by using the full sensitivity of ALMA.
Based on our CH$_3$NH$_2$ observations we estimate the column density of glycine (conformer\,I) to be $1.1\times10^{15}$\,cm$^{-2}$ by using the calculated fractional abundance of $4.6\times10^{-9}$ which corresponds to the ``medium warm-up'' case \citep{Garr13}.
We adopted the ``medium warm-up'' case since it matches best with our observed abundances of CH$_2$NH and CH$_3$NH$_2$.
It is generally thought that molecules formed on the dust grain surface are evaporated by UV photons from a central star.
The evaporation temperature can be estimated by utilizing the Clausius-Clapeyron equation together with a theoretically calculated boiling point of $487\pm23$\,K (see ChemSpider http://www.chemspider.com/).
For the fractional abundance of glycine, $4.6\times10^{-9}$, and a typical number density of H$_2$ for a hot core of $10^7$~cm$^{-3}$ \citep{Cesa94}, the evaporation temperature is calculated to be between 82 and 90\,K, which is close to the evaporation temperature of water ice on the dust ($\sim 100$\,K).
This mean that the solid phase glycine would be evaporated along with the evaporation of water ice. 
Thus we assume that the excitation temperature of glycine is 100\,K.

\begin{figure}[h]
 \begin{center}
  \includegraphics[width=18cm]{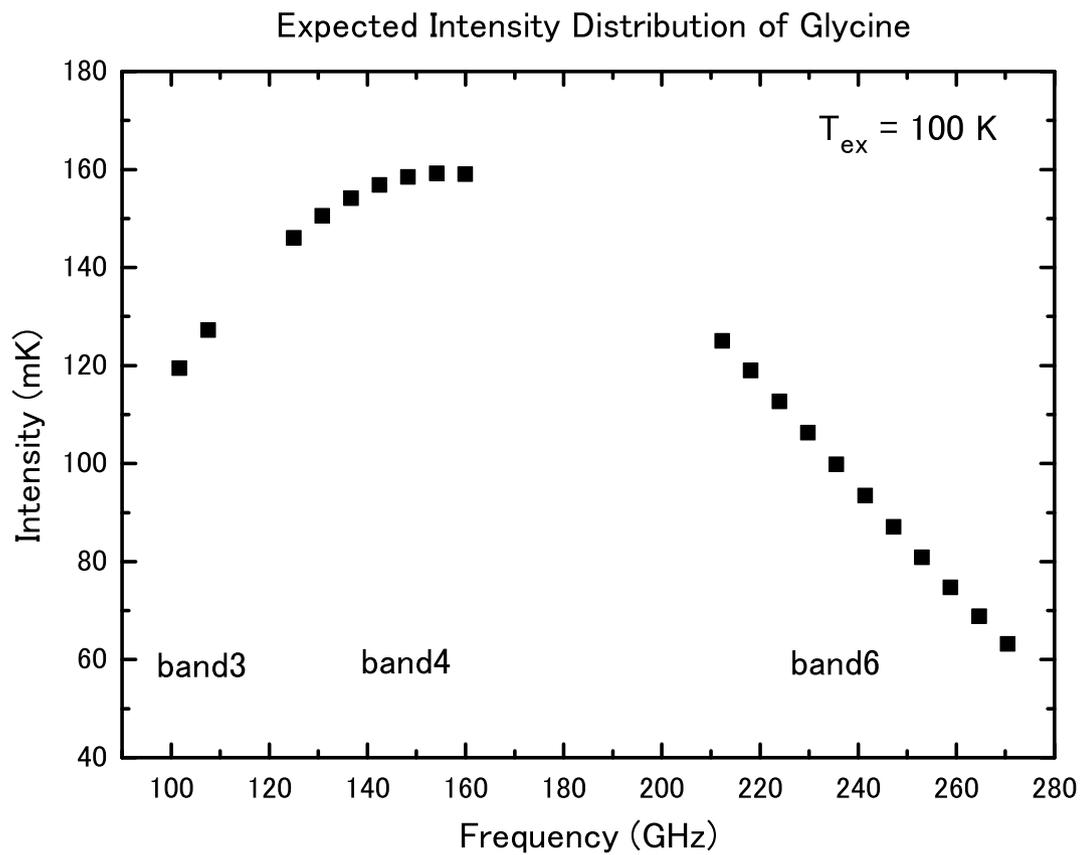} 
 \end{center}
\caption{An example of distribution of brightness temperature of CH$_3$NH$_2$ for the excitation temperature of 100\,K. The intensity calculation corresponds to the expected column density for G10.47+0.03. See the main text for the details. ALMA's band names are also shown.}\label{fig:exp_int}
\end{figure}

With an assumption of LTE, we calculated the brightness temperatures of glycine transitions.
Interestingly, four transitions (e.g., $J_{K, J-K}$ - $J-1_{K, J-K}$, $J_{K+1, J-K}$ - $J-1_{K+1, J-K}$, $J_{K, J-K}$ - $J-1_{K+1, J-K}$ and $J_{K+1, J-K}$ - $J-1_{K, J-K}$) are so close in frequency.
For K=0 or 1 transitions, their frequencies are within 300\,kHz at most, and they degenerate to form a single line.
In this way we co-added brightness temperatures of four closely located transitions, which are shown in Figure\,\ref{fig:exp_int}.
In Figure~\ref{fig:exp_int} we can easily see that the peak brightness temperatures are about 160\,mK in the frequency range between 150 and 160\,GHz.
For a case of higher excitation temperature of 150\,K, the frequency with peak brightness temperature will be shifted towards the higher frequencies, however, the calculated brightness temperatures between 150 and 160\,GHz still remain above 120\,mK.

A source size is an important parameter in discussing the detectability of molecular lines.
Assuming that glycine as well as water evaporate at 100\,K, the source size may correspond to a region with temperature of 100\,K and above.
The size can be estimated by using a methodology developed by \citet{Biss07} as a function of total luminosity of a source.
Then the estimated angular source size will be about 2 arcseconds.
This angular size corresponds to a linear size of 0.1\,pc at the distance of G10.47+0.03, which is comparable with a typical size of a hot molecular core.
In this case the detectability by using a single dish radio telescope would be very small; for the IRAM\,30\,m telescope the beam filling factor in the frequency range between 150 and 160\,GHz will be (2/20)$^{2}$=0.01.
Then beam-averaged brightness temperatures will drop down to 1.2-1.6\,mK, indicating that it is quite hard to detect glycine by single dish telescopes.
Therefore we need to utilize an interferometer, such as ALMA.
When we use the ALMA's Observing Tool (OT), we can estimate on-source integration times for detecting glycine; 5\,$\sigma$ detection of glycine lines can be achieved for on-source integration times of between 0.5 and 1.3 hours per line by using the full capability of ALMA.

\section{Conclusion}
We succeeded to find a new CH$_3$NH$_2$ source, G10.47+0.03, with its fractional abundance of $2.4\pm0.6\times10^{-8}$; at the time of writing, this source is the most abundant source of CH$_3$NH$_2$ ever known.
We also found another new source of CH$_3$NH$_2$, NGC6334F, however, we think the finding should be regarded tentative, considering the weakness of the signals.
We were not able to detect CH$_3$NH$_2$ towards two other hot core sources, G31.41+0.31 and W51\,e1/e2.
Detailed analysis of the detected data revealed the detection of the hyperfine components of CH$_3$NH$_2$ for the first time.
We found that the observed abundance of CH$_3$NH$_2$ agrees fairly well with the theoretically predicted value by \citet{Garr13}.
Since CH$_3$NH$_2$ is regarded a plausible precursor to glycine, we discussed detectability of interstellar glycine.
We found that it is highly possible to detect interstellar glycine towards G10.47+0.03 by using ALMA, however, we also found it would be very difficult to detect interstellar glycine by single dish radio telescopes.

Towards the future it is suggested to conduct survey observations of CH$_3$NH$_2$ not only in hot core/ hot corino sources but in protoplanetary disks and comets in studying incorporation of interstellar prebiotic molecules such as glycine into planetary forming regions.

\begin{ack}
We are grateful all the staff members of Nobeyama Radio Observatory, the National Astronomical 
Observatory of Japan (NAOJ), for their support throughout our observations.
We thank Dr. Kaori Kobayashi of Toyama University, Japan, in discussing the laboratory measurements on the hyperfine structure lines of CH$_3$NH$_2$.
Our research was supported by the Department of Astrobiology, Center for Novel Science Initiatives, National Institutes of Natural Sciences (NINS) and by the JSPS Kakenhi Grant Numbers JP14J03618 and JP15H03646.
A part of the data analysis was made at the Astronomy Data Center, NAOJ. 
We utilized the Japanese Virtual Observatory (JVO; http://jvo.nao.ac.jp/) in finding relevant astronomical data.
This research has made use of NASA's Astrophysics Data System.
\end{ack}

%\appendix 
%\section*{Case of single paragraph}
%
%\section{Case of two or paragraphs}
%
%\section{Case of two or paragraphs}

%%%
% See the manual for the detail.
%%%

\newpage

\newpage

\newpage

\newpage

\newpage
\begin{center}
\begin{table}[h]
%  \caption{List of Observed Sources}       
  \tbl{List of Observed Sources}{%
    \begin{tabular}{cccccc}
  \hline \hline
  Name & RA (J2000) & Dec (J2000) & $V_{\rm LSR}$ (km\,s$^{-1}$) & Distance (kpc) & Ref.\\ 
  \hline
  NGC6334F	& \timeform{17h20m53.4s}	& \timeform{-35D47'01.0''}	& -7 	& 1.4 & 1\\
  G10.47+0.03	& \timeform{18h08m38.1s}	& \timeform{-19D51'49.4''}	& 68 & 8.6 & 2\\
  G31.41+0.31	& \timeform{18h47m34.6s}	& \timeform{-01D12'43.0''}	& 97 & 7.9 & 3\\
  W51 e1/e2	& \timeform{19h23m43.77s}	& \timeform{+14D30'25.9''}	& 57 & 5.4 & 4\\
  \hline
    \end{tabular}}\label{tab:obs}
    \begin{tabnote}
References: 1 \citet{Wu14}, 2 \citet{Sann14}, 3 \citet{Hern14}, 4 \citet{Sato10}
\end{tabnote}
\end{table}
\end{center}

\newpage

\begin{longtable}[h!]{lcrrr}
\caption{List of Observed CH$_3$NH$_2$ Lines}\\
\caption*{List of Observed CH$_3$NH$_2$ Lines}\\
\label{tab:lines} 
\\
%  \begin{center}
  \hline \hline
  Transition & $F$'-$F$'' & Frequency (MHz) & $E_l$ (K) & $\mu^2S$ (D$^2$)\\ 
  \hline
  \endhead
  \hline
  \endfoot
  2$_1$\,E1-1 - 2$_0$\,E1+1	&  		& 78928.726	& 6.82	& 2.914 \\
                                        & 1-2 	& 78927.460 	& 6.82 	& 0.146 \\
                                        & 3-2	& 78927.927	& 6.82	& 0.151 \\
                                        & 1-1	& 78928.683	& 6.82	& 0.437 \\
                                        & 3-3	& 78928.713	& 6.82	& 1.209 \\
                                        & 2-2	& 78928.768	& 6.82	& 0.675 \\
                                        & 2-3	& 78929.554	& 6.82	& 0.151 \\
                                        & 2-1	& 78929.991	& 6.82 	& 0.146 \\
  1$_1$\,B1 - 1$_0$\,B2 		&  		& 79008.693 	& 2.42 	& 2.373 \\
                                        & 0-1	& 79007.177	& 2.42	& 0.264 \\
                                        & 2-1	& 79007.998	& 2.42	& 0.330 \\
                                        & 1-1	& 79008.546	& 2.42	& 0.198 \\
                                        & 2-2	& 79008.722	& 2.42	& 0.989 \\
                                        & 1-2	& 79009.270	& 2.42	& 0.330 \\
                                        & 1-0	& 79010.356	& 2.42	& 0.264 \\
  1$_1$\,E1-1 - 1$_0$\,E1+1 	&  		& 79210.297 	& 2.56 	& 1.392 \\
                                        & 1-1	& 79209.523	& 2.56	& 0.116 \\
                                        & 2-1	& 79209.728	& 2.56	& 0.193 \\
                                        & 0-1	& 79210.036	& 2.56	& 0.155 \\
                                        & 1-2	& 79210.247	& 2.56	& 0.193 \\
                                        & 2-2	& 79210.452	& 2.56	& 0.580 \\
                                        & 1-0	& 79211.333	& 2.52	& 0.155 \\
  6$_1$\,A2 - 6$_0$\,A1 		&  		& 81521.078 	& 44.67	& 10.824 \\
                                        & 5-5	& 81520.662	& 44.67	& 2.968 \\
                                        & 7-7	& 81520.816	& 44.67	& 4.078 \\
                                        & 6-6	& 81521.731	& 44.67	& 3.438 \\
  5$_1$\,A1 - 5$_0$\,A2 		&  		& 83978.941 	& 31.91	& 9.024 \\
                                        & 4-4	& 83978.505	& 31.91	& 2.363 \\
                                        & 6-6	& 83978.690	& 31.91	& 3.456 \\
                                        & 5-5	& 83979.595	& 31.91	& 2.811 \\
  4$_1$\,A2 - 4$_0$\,A1 		&  		& 86074.729 	& 21.28	& 7.290 \\
                                        & 3-3	& 86074.273	& 21.28	& 1.772 \\
                                        & 5-5	& 86074.497	& 21.28	& 2.851 \\
                                        & 4-4	& 86075.367	& 21.28	& 2.193 \\
%  3$_1$ A1 - 3$_0$ A2 		&  & 87782.494 	& 12.77		& 5.613 \\
  \pagebreak
  2$_1$\,A2 - 2$_0$\,A1 		&  		& 89081.463 	& 6.38 	& 3.978 \\
                                        & 1-2	& 89079.791	& 6.38	& 0.199 \\
                                        & 3-2	& 89080.548	& 6.38	& 0.206 \\
                                        & 1-1	& 89081.014	& 6.38	& 0.597 \\
                                        & 3-3	& 89081.334	& 6.38	& 1.650 \\
                                        & 2-2	& 89081.911	& 6.38	& 0.921 \\
                                        & 2-3	& 89082.697	& 6.38	& 0.206 \\
                                        & 2-1	& 89083.134	& 6.38	& 0.199 \\
  \hline
%  \end{center}
\end{longtable}
\begin{tabnote}
Transition parameters are taken from \citet{Ilyu07}. The quantum number, $F$, denotes the total rotational angular momentum including the hyper-fine splitting due to the quadrupole moment ($I$=1) of the $^{14}{\rm N}$ atom. $F$' and $F$'', respectively, correspond to the upper and lower levels. Transition frequencies without $F$ numbers correspond to those without considering the hyper-fine splittings due to the quadrupole moment ($I$=1) of the $^{14}{\rm N}$ atom.
\end{tabnote}

\newpage
\begin{table}[h]
  \caption{Gaussian-Fit Line Parameters of Observed CH$_3$NH$_2$ Lines towards G10.47+0.03}\label{tab:gauss_G10}       
  \begin{center}
    \begin{tabular}{cccccc}
  \hline \hline
  Transition & Rest Frequency & $T_{\rm A}^*$ & $V_{\rm LSR}$ & ${\Delta}v$ & rms\\ 
    & (MHz) & (mK) &  (kms$^{-1}$) &  (kms$^{-1}$) & (mK) \\
  \hline
  2$_1$\,E1-1 - 2$_0$\,E1+1		& 78928.713 	& 20		& 68.1 & 7.4 & 8 \\
  1$_1$\,B1 - 1$_0$\,B2 		& 79008.722 	& 25		& 68.3 & 5.4 & 8 \\
  1$_1$\,E1-1 - 1$_0$\,E1+1 	& 79210.452 	& 26		& 67.3 & 5.6 & 9 \\
  6$_1$\,A2 - 6$_0$\,A1 		& 81520.816 	& 25		& 67.9 & 5.4 & 8 \\
  5$_1$\,A1 - 5$_0$\,A2 		& 83978.941 	& $<24$ 	& -- 	  & --  & 8 \\
  4$_1$\,A2 - 4$_0$\,A1 		& 86074.497 	& 21		& 69.5 & 3.4 & 11 \\
%  3$_1$ A1 - 3$_0$ A2 		& 87782.494 	& 20		& 68 & 5 & 10 & 20 & -7 & 5 & 15 \\
  2$_1$\,A2 - 2$_0$\,A1 		& 89081.334 	& 27		& 68.2 & 2.8 & 8 \\
  \hline
    \end{tabular}
  \end{center}
\end{table}

\newpage
\begin{longtable}[c]{ccccccccc}
\caption{Gaussian-Fit Line Parameters of Possibly Observed CH$_3$NH$_2$ Lines towards NGC6334F}
\label{tab:gauss_N6334} 
\\
%  \begin{center}
%\small
  \hline \hline
  Transition & $F$'-$F$'' & Rest Freq. & Obs. Freq. & $\mu^2S$  & $T_{\rm A}^*$ & $V_{\rm LSR}$ & ${\Delta}v$ & rms\\ 
      & & (MHz) & (MHz) & (D$^2$) & (mK) & (kms$^{-1}$) &  (kms$^{-1}$) & (mK) \\
  \hline
  \endhead
  \hline
  \endfoot
  2$_1$\,E1-1 - 2$_0$\,E1+1 & 1-2	& 78927.460	&           & 0.146 & & & &\\
                            & 3-2	& 78927.927	&           & 0.151 & & & &\\
                            & 1-1	& 78928.683	&           & 0.437 & & & &\\
                            & 3-3	& 78928.713	& 78928.344 & 1.209 & 20 & -5.6 & 2.6 & 6 \\
                            & 2-2	& 78928.768	&           & 0.675 & & & &\\
                            & 2-3	& 78929.554	&           & 0.151 & & & &\\
                            & 2-1	& 78929.991	&           & 0.146 & & & &\\
  1$_1$\,B1 - 1$_0$\,B2     & 0-1	& 79007.177	&           & 0.264 & & & &\\
                          & 2-1	& 79007.998	& 79007.930 & 0.330 & 17 & -6.7 & 3.3 & 6 \\
                          & 1-1	& 79008.546	&           & 0.198 & & & &\\
                          & 2-2	& 79008.722	& 79008.539 & 0.989 & 23 & -6.3 & 1.9 & 6 \\
                          & 1-2	& 79009.270	& 79009.602 & 0.330 & 14 & -8.3 & 1.8 & 6 \\
                          & 1-0	& 79010.356	& 79010.188 & 0.264 & 15 & -6.4 & 1.6 & 6 \\
  1$_1$\,E1-1 - 1$_0$\,E1+1 & 1-1	& 79209.523	&           & 0.116 & & & &\\
                          & 2-1	& 79209.728	& 79209.344 & 0.193 & 11 & -5.6 & 3.9 & 5 \\
                          & 0-1	& 79210.036	&           & 0.155 & & & &\\
                          & 1-2	& 79210.247	&           & 0.193 & & & &\\
                          & 2-2	& 79210.452	& 79210.094 & 0.580 & 31 & -5.7 & 2.5 & 5 \\
                          & 1-0	& 79211.333	&           & 0.155 & & & &\\
  6$_1$\,A2 - 6$_0$\,A1     & 5-5	& 81520.662	&           & 2.968 & & & &\\
                          & 7-7	& 81520.816	& 81520.172 & 4.078 & 33 & -4.6 & 3.6 & 7 \\
                          & 6-6	& 81521.731	& 81521.250 & 3.438 & 23 & -5.2 & 2.3 & 7 \\
  5$_1$\,A1 - 5$_0$\,A2     & 4-4	& 83978.505	&           & 2.363 & & & &\\
                          & 6-6	& 83978.690	& 83979.227 & 3.456 & 21 & -8.9 & 2.2 & 4 \\
                          & 5-5	& 83979.595	& 83980.445 & 2.811 & 21 & -10.0 & 1.7 & 4 \\
%  4$_1$ A2 - 4$_0$ A1     & 3-3	& 86074.273	& & 1.772 \\
%                          & 5-5	& 86074.497	& & 2.851 \\
%                          & 4-4	& 86075.367	& & 2.193 \\
%  \pagebreak
  2$_1$\,A2 - 2$_0$\,A1     & 1-2	& 89079.791	&           & 0.199 & & & &\\
                          & 3-2	& 89080.548	&           & 0.206 & & & &\\
                          & 1-1	& 89081.014	&           & 0.597 & & & &\\
                          & 3-3	& 89081.334	& 89081.742 & 1.650 & 34 & -8.4 & 3.4 & 7\\
                          & 2-2	& 89081.911	&           & 0.921 & & & &\\
                          & 2-3	& 89082.697	&           & 0.206 & & & &\\
                          & 2-1	& 89083.134	&           & 0.199 & & & &\\
  \hline
%\normal
%  \end{center}
\end{longtable}

\newpage

\begin{table}[h]
  \caption{Comparison between Simulated and Observed Fractional Abundances}\label{tab:comparison}       
  \begin{center}
    \begin{tabular}{lccccc}
    \hline \hline
     & \multicolumn{3}{c}{Simulated Values$^{\rm a}$} & \multicolumn{2}{c}{Observed Values$^{\rm b}$} \\
  \hline
  Species & Fast & Medium & Slow & G10.47+0.03 & NGC6334F \\
  \hline
  CH$_2$NH & $1.1\times10^{-8}$ & $2.4\times10^{-8}$ & $1.5\times10^{-8}$ & $7\times10^{-8}$$^{\rm c}$ & $3\times10^{-9}$$^{\rm c}$ \\
  CH$_3$NH$_2$ & $8.0\times10^{-8}$& $3.6\times10^{-8}$ & $6.8\times10^{-9}$ & $1.2\times10^{-8}$$^{\rm d}$ & $2.5\times10^{-10}$$^{\rm d}$ \\
  NH$_2$CH$_2$COOH & $8.4\times10^{-11}$ & $4.6\times10^{-10}$ & $8.1\times10^{-9}$ & --- & --- \\
  \hline
    \end{tabular}
\begin{tabnote}
Notes: a -- peak values taken from Table\,8 of \citet{Garr13}; b -- The observed abundances are divided by a factor of 2 in order to align with the simulated values; c -- taken from Table\,5 of \citet{Suzu16}; d -- this work.
\end{tabnote}
\end{center}
\end{table}

\end{document}